\documentclass[conference]{IEEEtran}
\IEEEoverridecommandlockouts
\usepackage{cite}
\usepackage{amsmath,amssymb,amsfonts}
\usepackage{algorithmic}
\usepackage{graphicx}
\usepackage{textcomp}
\usepackage{xcolor}
\usepackage{booktabs}
\def\BibTeX{{\rm B\kern-.05em{\sc i\kern-.025em b}\kern-.08em
    T\kern-.1667em\lower.7ex\hbox{E}\kern-.125emX}}
\usepackage[utf8]{inputenc}
\usepackage{graphicx}
\usepackage[ruled,vlined,linesnumbered]{algorithm2e}

\title{Quantifying the Efficacy of Logic Locking Methods}
\date{August 2020}

\begin{document}

\author{\IEEEauthorblockN{Joseph Sweeney, Deepali Garg, Lawrence Pileggi}
\IEEEauthorblockA{\textit{Dept. of Electrical and Computer Engineering} \\
\textit{Carnegie Mellon University}\\
Pittsburgh, Pennsylvania \\
\{joesweeney,deepalig,pileggi\}@cmu.edu}
}

\maketitle

\begin{abstract}
The outsourced manufacturing of integrated circuits has increased the risk of intellectual property theft. In response, logic locking techniques have been developed for protecting designs by adding programmable elements to the circuit. These techniques differ significantly in both overhead and resistance to various attacks, leaving designers unable to discern their efficacy. To overcome this critical impediment for the adoption of logic locking, we propose two metrics, key corruption and minimum corruption, that capture the goals of locking under different attack scenarios. We develop a flow for approximating these metrics on generic locked circuits and evaluate several locking techniques.
\end{abstract}


\section{Introduction}
Due to prohibitively high research and development costs, only a few foundries are manufacturing integrated circuits (ICs) in advanced technology nodes. Consequently, many IC companies tend to operate fabless, relying on untrusted foundries to manufacture their designs. Once a circuit is sent for fabrication, the foundry gains full visibility of the design in the netlist and layout form, thereby enabling possible intellectual property (IP) theft.  This threat has two main forms: \textit{overproduction} in which ICs are manufactured without the designer's consent and \textit{leakage} in which the design itself or sensitive details are compromised. 
This threat undermines the significant cost associated with developing digital circuits and is a growing concern in the IC industry \cite{6860363}.

To combat IP theft, a variety of logic locking techniques have been developed. Generally, these techniques add programmable elements to the logic of an IC. When programmed incorrectly, the elements disrupt the circuit's functionality. 
The key, which correctly programs the elements, is set post-manufacture using a tamper-proof memory, so the correct design is never revealed to the untrusted entity. 
Corresponding with the logic locking schemes, powerful attacks have been developed that can reveal the circuit's correct key. 
Publishing in this area has risen dramatically as locks are proposed, broken, fixed, and again broken\cite{tan2020benchmarking}. 
This has led to a multitude of locking schemes, many of which are conceptually similar locks with slightly different implementation details. 
Lack of well-defined metrics and inconsistent attack assumptions makes the comparison of techniques difficult. 
Some locks have been written off as broken only to be replaced by more costly locks that provide worse security under some attack models. 

In this paper, we develop metrics that capture the notion of security under two common attack models. We then use these metrics to evaluate several different locking schemes. As compared to the conventional analyses, these metrics allow for more nuanced insights into the security that locks provide. Furthermore, the metrics enable a comparison of the overhead-security trade-offs associated with a given technique, \textit{an essential requirement for bringing logic locking into real systems}.  
Specifically, the contributions of this work are the following:
\begin{itemize}
\item Overview and open-source implementation of several logic locking classes and corresponding attack methods
\item Proposal of minimum corruption and key corruption metrics
\item Demonstration of scalable methods to approximate the proposed metrics and application of metrics to various attack scenarios
\item Analysis of the overhead versus corruption of locking techniques
\end{itemize}

\section{Background}
\label{sec:tax}
\subsection{Attack Models}
\label{sec:attack_models}
The vast majority of locking procedures can be described by the following model. A circuit is defined as a Boolean function, $C : X \rightarrow Y$, mapping an input space $X=\{0,1\}^n$ to an output space $Y=\{0,1\}^m$.
The circuit is locked via a transform that adds key inputs to the circuit, $\mathcal{L}: C\to C_l$, where $C_l : X \times K \rightarrow Y$ and $K=\{0,1\}^w$. 
When the correct key, $k_c$, is applied, the locked circuit produces the same input-output behavior as the original circuit, $\forall x , C_l(x,k_c)=C(x)$. 
Ideally, under other keys, the locked circuit behaves incorrectly to the extent that it is unusable.

The most rigorous definition of a successful attack is finding a functionally equivalent key. Here, the problem solved by the attacker is one of finding an exact key, $k_e$, for which the locked circuit produces the same function as the original, $k_e : \forall x \, C(x,k_e)= C(x,k_c)$. A relaxed version of the exact recovery success criteria is approximate recovery. Here, the attacker finds an approximate key, $k_a$, under which functionality of the locked circuit differs from the correct functionality with at most some probability $\epsilon$, formally, $k_a : P_{x \in X}[C_l(x,k_a) \neq C_l(x,k_c)] <\epsilon$. 

There are generally three attack scenarios considered in locking a circuit, differing in access to artifacts and abilities. 

\subsubsection{Netlist}
Adversarial access to the design data is a basic assumption of logic locking. Analyzing this data can produce a netlist containing the design's standard cells and interconnections. Along with the netlist, an adversary may have knowledge of transistor and interconnect models, allowing detailed physical simulation and analysis. 

\subsubsection{Oracle}
A more powerful attack model assumes access to a functioning version of the unlocked design. The unlocked circuit has the correct key set in its tamper-proof memory, affording the attacker black-box access, commonly referred to as an oracle. Obtaining an oracle may be trivial if the IC is available on the open market, but also could be the result of compromised physical security. 
In this paper and commonly in the field, it is assumed that the adversary has access to the unlocked design's scan chains. This allows an adversary to consider the circuit with only a combinational model. 

\subsubsection{Probe}
The use of probing techniques may provide means to directly reveal key bits from the oracle. 
Foundries commonly use probing to aid the development of manufacturing processes and circuit failure analysis. 
Key values have been probed in nodes as small as 28nm \cite{keymat}. However, as feature sizes continue to scale, this probing becomes more difficult, limited by spatial resolution \cite{tan2020benchmarking}. 
Most work in logic locking has not directly considered this threat, relying on the validity of the tamper-proof memory assumption. 

\subsection{Insertion Locks}
\label{sec:xor}
The earliest logic locking techniques insert parity gates (XOR/XNOR) into the circuit structure to invert nets depending on a key input\cite{J.A.Roy2008EPIC:Circuits}. 
The parity gates, termed key gates, are spliced into nets selected randomly or with heuristics maximizing corruption\cite{Rajendran2015FaultEncryption}. The circuit is subsequently resynthesized. 
These manipulations create a large amount of corruption within the circuit as the parity gate's entire upstream function is inverted. 
Similar techniques utilize multiplexors (MUXs) and lookup tables (LUTs) in lieu of parity gates\cite{Kamali2018LUTLockAN,6616532}. 
Importantly, these techniques exhibit low overhead as few additional gates are added. These methods were originally developed targeting an oracle attack model. Unfortunately, within this context, these methods have been largely broken using a miter-based satisfiability (SAT) attack, detailed in section \ref{mbsa}.

\subsection{Miter-Based SAT Attack}
\label{mbsa}
First proposed in \cite{Subramanyan2015EvaluatingAlgorithms}, this attack uses a SAT solver, the locked netlist and an unlocked circuit to iteratively produce input-output (IO) relationships. These relationships are used to rule out all keys that do not produce the same behavior, narrowing the space of possible circuit functionalities. The IO relationships are efficiently learned through a three-step procedure: \textbf{1.} First, a miter circuit, $M_0\equiv C_l(x,k_0)\neq C_l(x,k_1)$, is encoded into a SAT solver to determine an input that produces different behavior for two different keys. \textbf{2.} Next, the produced input, $x$, is applied to the unlocked circuit to determine the output, forming an input-output (IO) pair, $(x_0,y_0)\equiv (x,C_l(x,k_c))$, which the correct key must respect. \textbf{3.} The IO pair is added as a constraint to the miter circuit for the next iteration, $M_i\equiv M_{i-1} \wedge (C_l(x_{i-1},k_0)= y_{i-1})\wedge (C_l(x_{i-1},k_1)= y_{i-1})$. Now, any keys that satisfy the miter circuit will also satisfy the learned IO relationship, ruling out at least one of the keys from step 1. These steps repeat, adding more constraints until the miter circuit is unsatisfiable. At this point, any key that respects all learned IO relationships will be functionally correct. 

\subsection{Point Function-Based and Densely-Interconnected Locks}
Following the miter-based SAT attack, the authors in \cite{Subramanyan2015EvaluatingAlgorithms} discuss a 
point function-like structure formed from an
AND-tree locked with parity gates, $\wedge_{i=1}^n x^i\oplus k^i$, where $x^i$ is the $i$th bit of $x$. The structure requires an exponential \textit{number of iterations} under their algorithm. 
This observation led to a series of locking schemes that incorporate close relatives of this structure to resist the miter-based SAT attacks.
To further resist removal attacks based on analysis of properties such as signal probability \cite{Yasin2017SecurityAnti-SATb} and Boolean sensitivity \cite{sensitivity}, many versions of this locking scheme have been proposed. 
While the exact scaling of the SAT-resistance depends on the specific technique, the class shows the greatest miter-based SAT attack resistance when the number of incorrect input values is minimal. Thus, there is an inherent trade-off between attack resistance and the corruption of the circuit. 

Another approach to resisting the miter-based SAT attack is adding densely-interconnected instances into the circuit, overwhelming the SAT solver \cite{Kamali2019Full-Lock,Shamsi2018Cross-Lock:Architectures,modeling}. 
These instances typically have many interdependent keys. A prototypical example is LUTs combined with configurable routing, the resulting lock resembling a field-programmable gate array (FPGA) embedded into the circuit. The various locking methods vary on the specific insertion methods, density, and mixing with original logic. 

\section{Locking Metrics}
A pervasive problem in the logic locking community is the lack of metrics that adequately capture the intended notions of security. 
Additionally, the metrics that do exist are only evaluated on trivially small circuits or rely on closed-form equations that correspond to simple structured locking schemes. In this section, we describe the limitations of existing metrics. We then define two metrics that capture intuitive definitions of security for the netlist and oracle attack models and subsequently relate the metric evaluation problem to the well-established field of model counting. 

\subsection{Existing Security Metrics}
The run-time of the miter-based SAT attack has been a ubiquitous metric since the attack's debut. The typical demonstration sweeps the number of key bits and produces a (hopefully) exponentially scaling attack time. 
While resistance to this attack is essential if the oracle attack model is considered, this run-time may give an over-optimistic notion of security. Importantly, \textit{while running the attack to completion may be infeasible, the intermediate results may produce keys that are functionally close to the original design}. In this case, the adversary need not complete the attack, but rather run until the keys produced exhibit low enough error rates. 

\begin{figure}[t]
  \centering
  \includegraphics[width=.9\columnwidth]{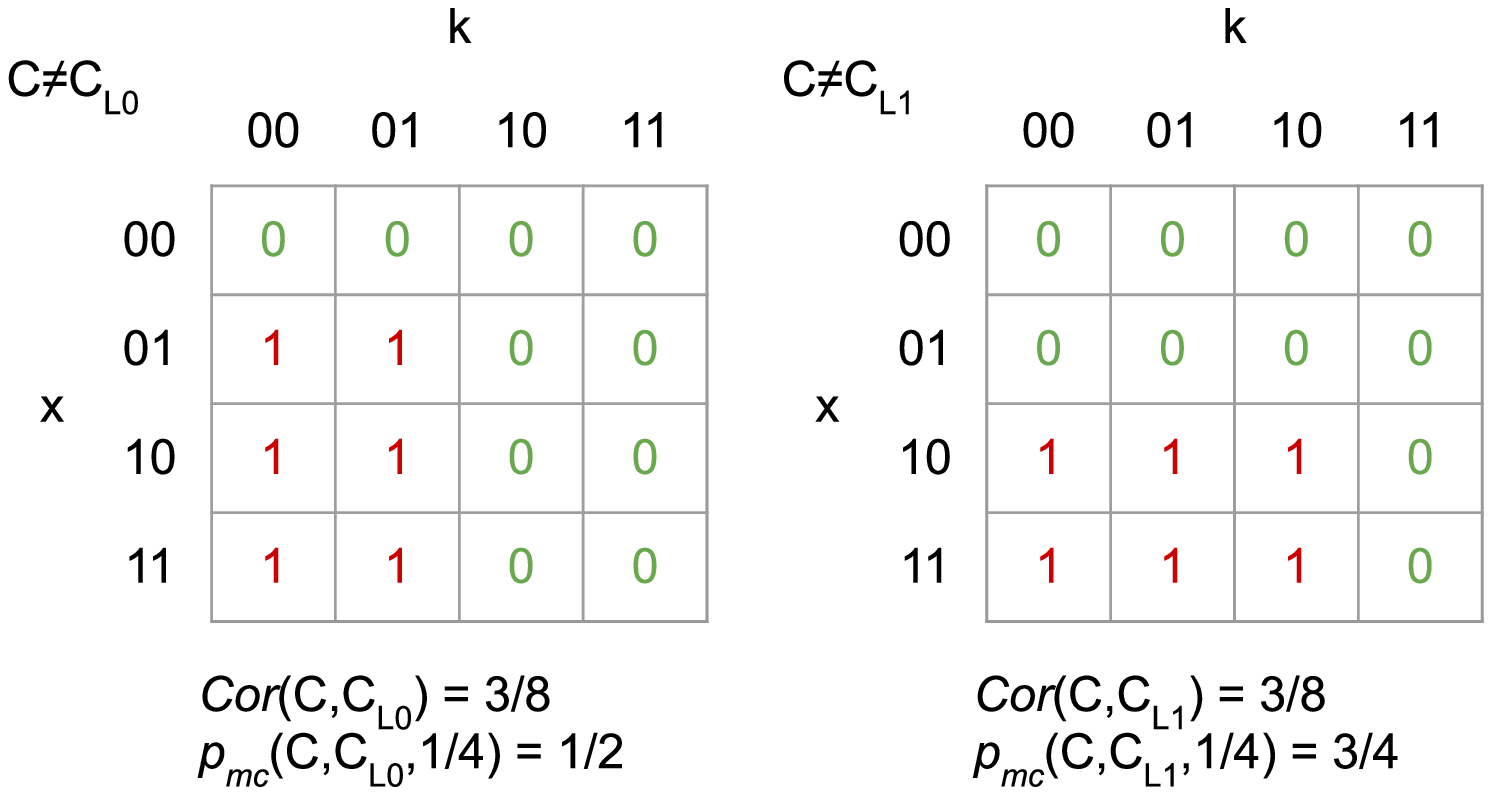}
  \caption{Miter truth tables and corresponding locking metrics for two locked circuits.}
  \label{fig:cor}
\end{figure}

A previously proposed metric for assessing a lock quality in terms of error rates is
corruptibility\cite{8395439}, defined as \[Cor(C,C_l) \equiv P_{x\in X, k \in K}[C(x) \neq C_l(x,k)] \] This metric captures the likelihood across all keys and inputs that a locked circuit is incorrect, essentially the total amount of inaccuracy in a locked circuit. However, it gives no notion of the distribution of incorrect values. 
An example of why this is important is depicted in Fig. \ref{fig:cor}. 
Here, a circuit is locked with two different schemes that produce locked circuits $C_{L0}$ and $C_{L1}$.
The resulting miter truth tables, $C\neq C_{L0}$ and $C\neq C_{L1}$ are shown. 
Both locked circuits exhibit the same corruptibility, however, the quality of the locking schemes is clearly different. For half of the possible keys, $C_{L0}$ is completely correct whereas $C_{L1}$ has incorrect values for every key other than the correct one. This motivates the development of a metric that can capture this disparity.

\subsection{Key Corruption}
Our first metric, key corruption, is meant to capture a more precise notion of resistance to oracle-based attacks beyond run-time. 
Key corruption is the portion of the input space that is mapped to incorrect outputs for a given key. 
Specifically, it is defined as \[ KeyCor(C,C_l,k) \equiv P_{x\in X}[C(x) \neq C_l(x,k)] \] 
This metric directly corresponds to the approximate key recovery outlined in section \ref{sec:attack_models}. 
This is useful to the designer in 
assessing the accuracy of intermediate keys produced by an attack and also
as a kernel in computing more complex metrics. 

To evaluate the key corruption, we build a miter circuit, $M \equiv C(x)\neq C_l(x,k_\mathrm{attack})$. Counting the number of input values that satisfy this miter and normalizing by the size of the circuit's input space will determine the key corruption. Typical circuits can have input widths upwards of 64 bits, thus it is necessary to utilize approximation techniques as outlined in section \ref{sec:em}.
The error of the estimated count of satisfying solutions can be tied to the bounds of the approximate solver. 

Depending on the circuit's application, the targeted key corruption definition can be adapted. The above definition counts an input value as incorrect if it has at least one output bit incorrect. Some applications may require several output bits to be incorrect. In this case key corruption could be defined as $P_{x\in X}[(\sum_{i=1}^m C(x)^i \oplus C_l(x,k)^i)\geq t]$, where $t$ is the required number of incorrect bits. Other definitions could weight certain bits more heavily, potentially useful in ensuring significant error in arithmetic operations. 

\subsection{Minimum Corruption}
\label{sec:mincor}
While key corruption can assess the progress of oracle-based attacks, under a netlist attack model, a more useful metric to a designer is the probability that a sampled key meets a certain corruption threshold, $\epsilon\in [0,1]$. 
This threshold can be determined by the application. For example, disabling a cryptographic function may only require a small portion of the inputs to be corrupted whereas a neural network accelerator may need significantly more. 
We capture this notion using minimum corruption, defined as
\[ MinCor(C,C_l,k,\epsilon) \equiv KeyCor(k) \geq \epsilon \]
We then can define a probability of selecting a key that meets this minimum corruption value, $p_{mc}$.
\[ p_{mc}(C,C_l,\epsilon) \equiv P_{k \in K}[MinCor(C,C_l,k,\epsilon)] \]
For a given key, minimum corruption discounts the corruption beyond the threshold.
Again considering Fig. \ref{fig:cor}, we see that $p_{mc}$ captures the difference between the two locking scenarios. 
The designer can determine a suitable threshold, then can scale the amount of locking until the probability of meeting the corruption threshold is acceptable.  

The techniques discussed in section \ref{sec:em} can be used to efficiently approximate this metric. 
To obtain a $p_{mc}$ estimate, we take a set of uniform random samples from the key space.
For each of these sampled keys, we estimate the key corruption and compare it to the threshold to determine if it meets the minimum corruption. The fraction of key samples for which key corruption is greater than the threshold computes $p_{mc}$.

\subsection{Estimating Metrics}
\label{sec:em}
In general, the metrics we propose will be used to evaluate the amount of discrepancy between the original and locked circuits under various scenarios. 
Calculation of these metrics directly maps to model counting (\#SAT). 
The evaluation of metrics can be encoded as a Boolean formula wherein the number of satisfying solutions over the total space gives the value. Due to the high dimensionality of the problem, exact solutions can only be obtained for a limited key and input width. 
To understand how locks scale, we resort to approximation methods. 
Luckily, approximate model counting is a widely studied area with many efficient, open-source solvers. 

We use the solver ApproxMC\cite{CMV16} as a kernel in estimating our proposed metrics.
This solver uses hash functions to split a circuit's input space into small, countable partitions of roughly equal size. 
By counting a single partition and multiplying by the number of partitions, the tool can give an estimation of the number of solutions to a formula. 
Repeating this process allows increased confidence in the estimation. 
Conveniently, ApproxMC has a rigorous formulation of probably approximately correct (PAC) bounds. The relation between the real count, $N$, and the estimated count, $N_{est}$, is parameterized by $\delta\in (0,1]$ and $\epsilon_a>0$. Specifically, the relationship is $P[N/(1+\epsilon_a) \leq N_{est}\leq N(1+\epsilon_a)] \geq 1-\delta$.

\section{Application of Metrics}
The metrics proposed in the previous section can be used to evaluate the efficacy of locking techniques under the netlist and oracle attack models. 
We demonstrate this process using representative locks from the classes described in section \ref{sec:tax}. 
We implement the locks using the open-source python library \textit{circuitgraph}, which allows for easy manipulation of netlists. 
To allow for reproducibility, we share our lock and metric implementations in our repository that will be made available to the community upon acceptance. 
From the insertion-based lock types, we implement XOR, LUT, and MUX locking\cite{J.A.Roy2008EPIC:Circuits,Kamali2018LUTLockAN,6616532}. From the point-function based locks, we implement SFLL-Flex\cite{YasinProvably-SecurePractice}. 
Finally, from the densely-interconnected locks, we implement Full-Lock\cite{Kamali2019Full-Lock} and LEBL\cite{modeling}. 
Our miter-based attack implementation uses the SAT solver CaDiCaL\cite{Cadical}.
For the locking techniques that produce cyclic circuits, we use CycSAT\cite{Zhou2017CycSAT:Encryptions} to form acyclic key conditions so that the attack terminates correctly.  
Each of these techniques is used to lock circuits from the ISCAS 85 combinational benchmark set\cite{Brglez}. 

\subsection{Minimum Corruption under Netlist Attack Model}
\label{sec:structure}
\begin{figure}[t]
  \centering
  \includegraphics[width=0.9\columnwidth]{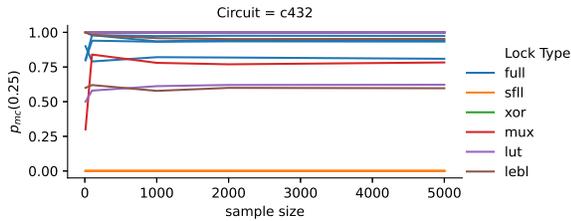}
  \caption{Approximate $p_{mc}$ of benchmark versus $\#(\mathrm{key samples})$.}
  \label{fig:saturation}
\end{figure}
\begin{figure}[t]
  \centering
  \includegraphics[width=0.9\columnwidth]{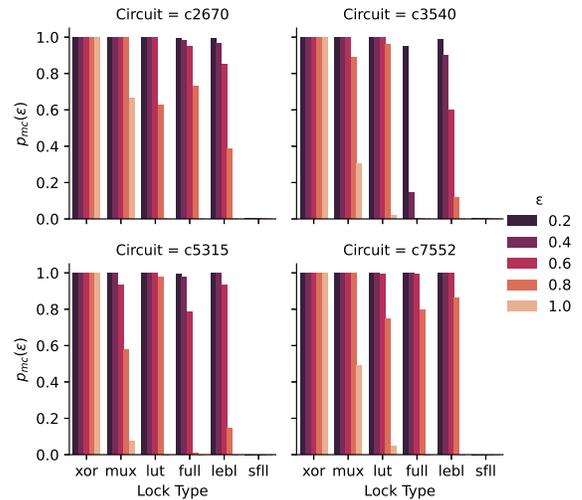}
  \caption{Approximate $p_{mc}$ of benchmark circuits locked with selected techniques. $w \approx $ 128 and $\#(\mathrm{key samples})=1000$}
  \label{fig:mincor}
\end{figure}
\begin{figure}[t]
  \centering
  \includegraphics[width=.8\columnwidth]{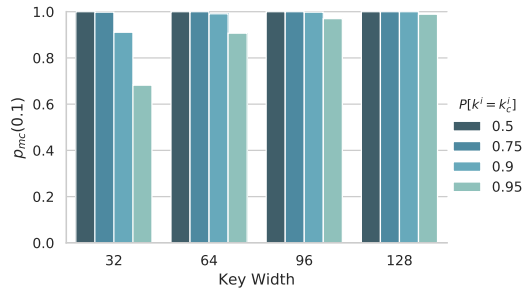}
  \caption{Approximate $p_{mc}$ of c3540 locked with XOR-locking, sweeping key width $w$ and $P[k^i=k^i_c]$ with $\#(\mathrm{key samples})=1000$.}
  \label{fig:biasmincor}
\end{figure}

As discussed in section \ref{sec:mincor}, the probability of meeting minimum corruption provides the designer with a good understanding of how likely it is for an adversary to select a key that functions close to the correct design. 
In Fig. \ref{fig:saturation}, we show an experimental run to determine an appropriate number of key samples to use for the estimation. As seen from the trend for our set of locked circuits, the value tends to converge around 1000 samples. We use this value for the remainder of our $p_{mc}$ estimates.

Assuming that the adversary has no a priori knowledge of the key bits, we can evaluate $p_{mc}$ for the selected locking techniques. 
We lock four circuits from our benchmark set with roughly 128 bits of locking. (The widths are not exact as the densely-interconnected techniques scale in unequal increments.) Using our estimation process, we uniformly sample keys and evaluate $p_{mc}$, showing the results in Fig. \ref{fig:mincor}. Several interesting conclusions can be drawn from this data. First, we see that XOR-locking shows the highest $p_{mc}$ value. At this key width, all keys sampled are corrupted for all inputs on at least one output. This shows the significant amount of corruption obtained from the inversion of random nets in the circuit. Considering the other techniques, we see that they all, except SFLL-Flex, have a high probability of meeting a 0.2 $\epsilon$ value. Generally, these probabilities decrease with $\epsilon$ at rates depending on the circuit and lock type. 

We can integrate more complex attacks into this analysis. For example, as discussed in \cite{8607163}, an adversary can analyze the local structure of a circuit locked with XOR-locking and determine a likely key with reported accuracy up to 95\%. The effect of this bias in the key space can be assessed with our minimum corruption metric. We assume that the designer requires at least 10\% of the input space to be corrupted. As the analysis of each locked gate is local, we assume that each key bit is independently drawn from a Bernoulli distribution with the probability parameter, $p$, set to the accuracy of the model, $P[k^i=k^i_c]\sim Bernoulli(p)$. We sweep $p$ from 0.5 (i.e. no information) to 0.95, the highest reported accuracy of the models. For each accuracy level, we determine the minimum corruption for a set of circuits with varying amounts of XOR-locking. 
As seen from the results in Fig. \ref{fig:biasmincor}, even at $p=0.95$ and 96 bits of locking, the value of $p_{mc}$ is very high. This shows that, while the local-structure analysis for likely key can significantly narrow the distribution of the correct keys, it does not necessarily translate into a circuit that is functionally close to the original; largely due to the high corruption of parity gates. 
If the attacker's goal is to produce a functionally correct or approximately correct circuit with the netlist alone, this attack scheme is unlikely to succeed. 

\subsection{Incremental Key Corruption of Oracle Attack Model}
\begin{figure}[t]
  \centering
  \includegraphics[width=\columnwidth]{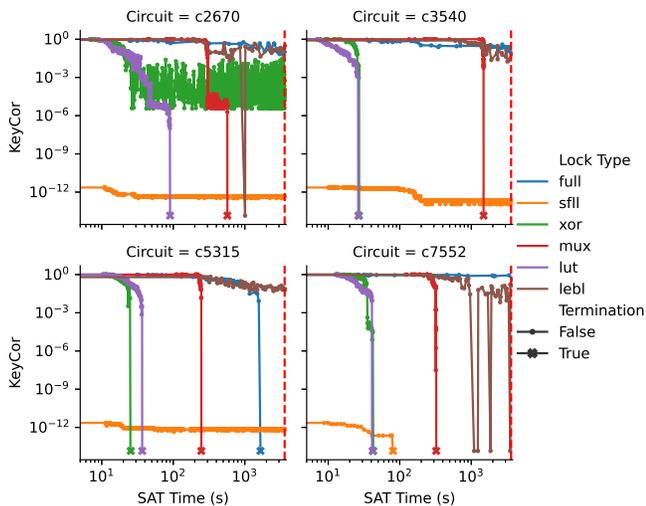}
  \caption{Key corruption for incremental keys returned from miter-based SAT attack, $w \approx $ 448. The attacks are run with a timeout of 1 hour, indicated by the dashed red line. The zero value of key corruption is mapped to $10^{-14}$}
  \label{fig:sat}
\end{figure}
The \textit{de facto} metric evaluated for oracle-based attacks is attack termination time. 
In this dimension, both the point function-based and densely-interconnected techniques exhibit very strong attack resistance.
However, these attack times mean little if unaccompanied by a notion of corruption for the remaining keys. 

Typically, when executing oracle-based attacks, a plausible key is produced in each iteration. Solving for additional keys is costly, likely motivating the attacker to simply use this incremental key. Evaluating the key corruption of the key from the attack can be used to indicate the progress of the attack. 

In Fig. \ref{fig:sat} we demonstrate the use of this metric to augment the miter-based SAT attack for our selected locking techniques.
We lock the circuits with roughly 448 key bits and run the attack with a timeout of 1 hour, evaluating the corruption at each iteration. 
The results show several interesting insights. First, we see that in most cases XOR and LUT-locking terminate in under 100 seconds. At such large key widths, it is clear that these techniques do not hold up under this attack model. MUX-locking takes about an order of magnitude longer to terminate. As expected, we see no termination in the SFLL results; however, the key corruption remains too low to likely have a significant effect. 
The densely-interconnected techniques Full-Lock and LEBL generally show the best results, with the highest corruption levels at the timeout. We do see one run terminating under an hour for Full-lock. \textit{The trend in the key corruption for LEBL reveals an interesting pattern undetectable by just considering the attack termination time.}
Several of the $\sim 100$ intermediate results for circuits c2670 and c7552 are functionally correct keys, a major security vulnerability. An astute attacker could thoroughly test the intermediate key results and confirm that these keys have arbitrarily low corruption with the oracle. 

\subsection{Overhead-Security Trade-Offs}
\begin{figure}[t]
  \centering
  \includegraphics[width=0.9\columnwidth]{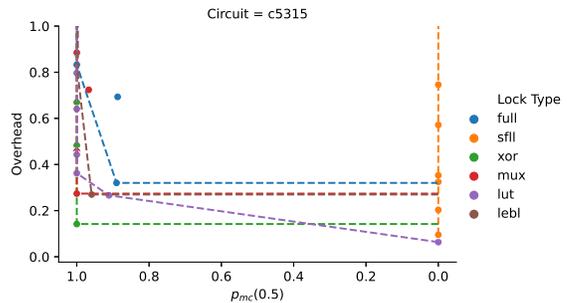}
  \caption{Overhead vs. $p_{mc}$ Pareto fronts for selected locking techniques.}
  \label{fig:net_ovr_cor}
\end{figure}
\begin{figure}[t]
  \centering
  \includegraphics[width=0.9\columnwidth]{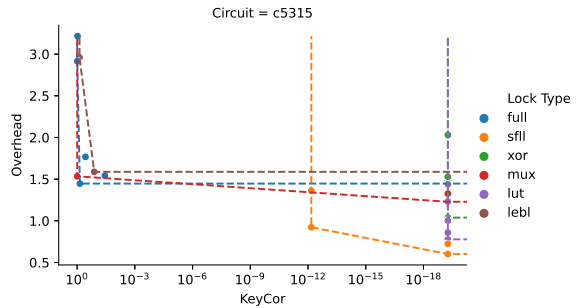}
  \caption{Overhead vs. key corruption Pareto fronts for selected locking techniques. The zero value of key corruption is mapped to $10^{-19}$}
  \label{fig:ora_ovr_cor}
\end{figure}

Overhead in the typical VLSI metrics, delay, area, and power is a critical concern of logic locking. The application of the IC may enforce limitations on the acceptable overhead. 
Even if this is not the case, too much overhead may motivate the use of commercial solutions such as FPGAs or microprocessors, rather than design an ASIC. We analyze our selected locking techniques across these metrics to show their scaling with the number of key bits.

Using Cadence Genus along with a commercial standard cell library in a 28nm process, we obtain overheads as follows. The maximum frequency of the design is found via iterative logic synthesis runs. This result serves as the baseline implementation to which various amounts of locking are applied and to which the results are normalized.
Each design is locked with the different techniques, varying key widths such that they produce roughly the same overhead range. We combine power, area, and delay into a single overhead value using, $Overhead \equiv (power_{locked}/power_{orig})\times(delay_{locked}/delay_{orig})\times(area_{locked}/area_{orig})-1$.
Coupling this data with an attack model and corresponding security metric, we can visualize the overhead-corruption trade-off. 

Fig. \ref{fig:net_ovr_cor} displays the trade-off under the netlist attack model.
We use $p_{mc}$ at $\epsilon=0.5$ plotted against the overhead. For each locking scheme, we draw the Pareto front. By a significant margin, the best performing locking scheme under these criteria is XOR-locking. With overhead less than 20\%, it shows a $p_{mc}(0.5) \approx 1$. 

The oracle attack model results are plotted in Fig. \ref{fig:ora_ovr_cor}. To handle the oscillations, we use the average $KeyCor$ of the keys produced in the last 100 seconds of the attack. 
Full-Lock shows the best result, closely followed by MUX-locking and LEBL. The other techniques appear as vertical lines since they terminated for all runs. Even though in Fig. \ref{fig:sat}, MUX-Locking has a lower per-bit key corruption, its low overhead makes it comperable to Full-Lock and LEBL.

\section{Discussion and Future Work}
The proposed metrics capture a more nuanced picture of a locking scheme than previous evaluation methods. 
By tailoring our metrics to specific scenarios, we can better understand what the adversary will be able to achieve along with the overhead costs. 
Additionally, our metrics provide a framework that can allow more detailed attack comparisons, as demonstrated with our analysis of XOR-locking under a structural analysis attack. 

Notably, our results capture the disparity between the netlist and oracle attack models. Under the netlist attacks, it is reasonable to expect the sampled keys will exhibit significant corruption. However, once an oracle is available, the corruption values drop by many orders of magnitude, even when substantially scaling the allowed overheads. This suggests that \textit{finding ways to prevent oracle access should be a topmost priority for the logic locking community}. 

Several interesting directions exist for future work. 
Because the vast majority of locking techniques simply modify the next state logic, we have evaluated these metrics in a combinational setting. As more sequential locking techniques are developed, extending the metrics to capture the sequential setting will be valuable. 
This can be achieved by unrolling the circuits as is done for sequential ATPG methods. 

Another valuable extension within the oracle attack model is an estimation of required attack time to achieve a key corruption value. Trends in the key corruption over attack time could be used to extrapolate such a metric. 
Finally, while we have shown that these metrics can be evaluated for circuits with several thousands of gates, within the range of the logic cone sizes of many industrial circuits, understanding the limits of these approximation techniques would be useful. 

\section{Conclusion}
In this work, we have proposed two metrics for logic locking that capture notions of security for commonly considered attack models. We have shown how to estimate these metrics utilizing approximate model counting techniques. 
In evaluating the metrics on three families of locking techniques, we have shown previously unknown overhead-security trade-offs and vulnerabilities. 
Not only are these metrics critical information in the application of logic locking, but they can also be used to set quantitative goals for future locking schemes.

 \section*{Acknowledgment}
This work was supported in part by the Defense Advanced Research Projects Agency under contract FA8750-17-1-0059 “Obfuscated
Manufacturing for GPS (OMG)” and Honeywell Federal Manufacturing \& Technologies, LLC under contract A023646.

\bibliographystyle{IEEEtran}
{\small
\bibliography{ref}}

\begin{thebibliography}{10}
\providecommand{\url}[1]{#1}
\csname url@samestyle\endcsname
\providecommand{\newblock}{\relax}
\providecommand{\bibinfo}[2]{#2}
\providecommand{\BIBentrySTDinterwordspacing}{\spaceskip=0pt\relax}
\providecommand{\BIBentryALTinterwordstretchfactor}{4}
\providecommand{\BIBentryALTinterwordspacing}{\spaceskip=\fontdimen2\font plus
\BIBentryALTinterwordstretchfactor\fontdimen3\font minus
  \fontdimen4\font\relax}
\providecommand{\BIBforeignlanguage}[2]{{%
\expandafter\ifx\csname l@#1\endcsname\relax
\typeout{** WARNING: IEEEtran.bst: No hyphenation pattern has been}%
\typeout{** loaded for the language `#1'. Using the pattern for}%
\typeout{** the default language instead.}%
\else
\language=\csname l@#1\endcsname
\fi
#2}}
\providecommand{\BIBdecl}{\relax}
\BIBdecl

\bibitem{6860363}
M.~{Rostami}, F.~{Koushanfar}, and R.~{Karri}, ``A primer on hardware security:
  Models, methods, and metrics,'' \emph{Proceedings of the IEEE}, vol. 102,
  no.~8, pp. 1283--1295, 2014.

\bibitem{tan2020benchmarking}
B.~Tan, R.~Karri, N.~Limaye, A.~Sengupta, O.~Sinanoglu, M.~M. Rahman,
  S.~Bhunia, D.~Duvalsaint, R.~D., Blanton, A.~Rezaei, Y.~Shen, H.~Zhou, L.~Li,
  A.~Orailoglu, Z.~Han, A.~Benedetti, L.~Brignone, M.~Yasin, J.~Rajendran,
  M.~Zuzak, A.~Srivastava, U.~Guin, C.~Karfa, K.~Basu, V.~V. Menon, M.~French,
  P.~Song, F.~Stellari, G.-J. Nam, P.~Gadfort, A.~Althoff, J.~Tostenrude,
  S.~Fazzari, E.~Breckenfeld, and K.~Plaks, ``Benchmarking at the frontier of
  hardware security: Lessons from logic locking,'' 2020.

\bibitem{keymat}
M.~T. {Rahman}, S.~{Tajik}, M.~S. {Rahman}, M.~{Tehranipoor}, and
  N.~{Asadizanjani}, ``The key is left under the mat: On the inappropriate
  security assumption of logic locking schemes,'' \emph{Cryptology ePrint
  Archive, Report 2019/719, 2019, https://eprint.iacr.org/2019/719, to appear
  at HOST 2020.}

\bibitem{J.A.Roy2008EPIC:Circuits}
F.~K. J.~A.~Roy and I.~L. Markov, ``{EPIC: Ending Piracy of Integrated
  Circuits},'' \emph{2008 Design, Automation and Test in Europe}, 2008.

\bibitem{Rajendran2015FaultEncryption}
J.~Rajendran, H.~Zhang, C.~Zhang, G.~S. Rose, Y.~Pino, O.~Sinanoglu, and
  R.~Karri, ``{Fault Analysis-Based Logic Encryption},'' \emph{IEEE
  Transactions on Computers}, vol.~64, no.~2, pp. 410--424, 2015.

\bibitem{Kamali2018LUTLockAN}
H.~M. Kamali, K.~Z. Azar, K.~Gaj, H.~Homayoun, and A.~Sasan, ``Lut-lock: A
  novel lut-based logic obfuscation for fpga-bitstream and asic-hardware
  protection,'' \emph{2018 IEEE Computer Society Annual Symposium on VLSI
  (ISVLSI)}, pp. 405--410, 2018.

\bibitem{6616532}
J.~{Rajendran}, H.~{Zhang}, C.~{Zhang}, G.~S. {Rose}, Y.~{Pino},
  O.~{Sinanoglu}, and R.~{Karri}, ``Fault analysis-based logic encryption,''
  \emph{IEEE Transactions on Computers}, vol.~64, no.~2, pp. 410--424, 2015.

\bibitem{Subramanyan2015EvaluatingAlgorithms}
P.~Subramanyan, S.~Ray, and S.~Malik, ``{Evaluating the security of logic
  encryption algorithms},'' \emph{Proceedings of the 2015 IEEE International
  Symposium on Hardware-Oriented Security and Trust, HOST 2015}, pp. 137--143,
  2015.

\bibitem{Yasin2017SecurityAnti-SATb}
M.~Yasin, B.~Mazumdar, O.~Sinanoglu, and J.~Rajendran, ``{Security Analysis of
  Anti-SAT},'' in \emph{Asia and South Pacific Design Automation Conference
  (ASP-DAC)}.\hskip 1em plus 0.5em minus 0.4em\relax IEEE, 2017.

\bibitem{sensitivity}
J.~Sweeney, M.~J.~H. Heule, and L.~Pileggi, ``Sensitivity analysis of locked
  circuits,'' in \emph{LPAR23. LPAR-23: 23rd International Conference on Logic
  for Programming, Artificial Intelligence and Reasoning}, ser. EPiC Series in
  Computing, E.~Albert and L.~Kovacs, Eds., vol.~73.\hskip 1em plus 0.5em minus
  0.4em\relax EasyChair, 2020, pp. 483--497.

\bibitem{Kamali2019Full-Lock}
H.~M. Kamali, K.~Z. Azar, H.~Homayoun, and A.~Sasan, ``{Full-Lock}.''\hskip 1em
  plus 0.5em minus 0.4em\relax Association for Computing Machinery (ACM), 2019,
  pp. 1--6.

\bibitem{Shamsi2018Cross-Lock:Architectures}
\BIBentryALTinterwordspacing
K.~Shamsi, M.~Li, D.~Z. Pan, and Y.~Jin, ``{Cross-Lock: Dense Layout-Level
  Interconnect Locking using Cross-bar Architectures},'' 2018. [Online].
  Available: \url{https://doi.org/10.1145/3194554.}
\BIBentrySTDinterwordspacing

\bibitem{modeling}
J.~{Sweeney}, M.~J.~H. {Heule}, and L.~{Pileggi}, ``Modeling techniques for
  logic locking,'' \emph{International Conference on Computer Aided Design},
  2020.

\bibitem{8395439}
K.~{Shamsi}, T.~{Meade}, M.~{Li}, D.~Z. {Pan}, and Y.~{Jin}, ``On the
  approximation resiliency of logic locking and ic camouflaging schemes,''
  \emph{IEEE Transactions on Information Forensics and Security}, vol.~14,
  no.~2, pp. 347--359, 2019.

\bibitem{CMV16}
S.~Chakraborty, K.~S. Meel, and M.~Y. Vardi, ``Algorithmic improvements in
  approximate counting for probabilistic inference: From linear to logarithmic
  sat calls,'' in \emph{Proceedings of International Joint Conference on
  Artificial Intelligence (IJCAI)}, 7 2016.

\bibitem{YasinProvably-SecurePractice}
M.~Yasin, A.~Sengupta, M.~T. Nabeel, M.~Ashraf, J.~J. Rajendran, and
  O.~Sinanoglu, ``Provably-secure logic locking: From theory to practice,'' in
  \emph{Proceedings of the 2017 ACM SIGSAC Conference on Computer and
  Communications Security}, ser. CCS ’17.\hskip 1em plus 0.5em minus
  0.4em\relax New York, NY, USA: Association for Computing Machinery, 2017, p.
  1601–1618.

\bibitem{Cadical}
A.~Biere, ``Cadical, lingeling, plingeling, treengeling and yalsat entering the
  sat competition 2018,'' in \emph{Proceedings of SAT Competition 2018}, 2018,
  pp. 13--14.

\bibitem{Zhou2017CycSAT:Encryptions}
H.~Zhou, R.~Jiang, and S.~Kong, ``{CycSAT: SAT-based attack on cyclic logic
  encryptions},'' in \emph{IEEE/ACM International Conference on Computer-Aided
  Design, Digest of Technical Papers, ICCAD}, 2017.

\bibitem{Brglez}
F.~Brglez and H.~Fujiwara, ``A neutral netlist of 10 combinational benchmark
  circuits and a targeted translator in fortran,'' \emph{Special session on
  ATPG and fault simulation, Proc. IEEE International Symposium on Circuits and
  Systems}, pp. 663--698, 06 1985.

\bibitem{8607163}
P.~{Chakraborty}, J.~{Cruz}, and S.~{Bhunia}, ``Sail: Machine learning guided
  structural analysis attack on hardware obfuscation,'' in \emph{2018 Asian
  Hardware Oriented Security and Trust Symposium (AsianHOST)}, 2018, pp.
  56--61.

\end{thebibliography}

\end{document}